\documentclass[twoside,fleqn]{article}
\usepackage{espcrc2,amssymb,hyperref}
\usepackage{graphicx}

\title{Mesonic Form Factors}

\author{
  Frederic D.~R.~Bonnet,$^\mathrm{a,b}$
  Robert G.~Edwards,$^\mathrm{b}$
  George T.~Fleming,$^\mathrm{b}$
  Randy Lewis\address{
    Department of Physics,
    University of Regina,
    Regina, SK, S4S 0A2, Canada
  }
  and David G.~Richards\address{
    Thomas Jefferson National Accelerator Facility,
    Newport News, VA 23606, USA
  }
  \textsl{[Lattice Hadron Physics Collaboration]}\thanks{
    This work was supported in part by the Natural Sciences and Engineering
    Research Council of Canada and by the U.S. Department of Energy under
    contract DE-AC05-84ER40150.  Computations were performed on the 128-node
    Pentium IV cluster at JLab and on other resources at ORNL,
    under the auspices of the U.S.\ DoE's SciDAC initiative.
  }
}

\begin{document}

\pagestyle{empty}

\begin{abstract}

We have started a program to compute the electromagnetic form factors
of mesons.  We discuss the techniques used to compute the pion
form factor and present preliminary results computed with domain
wall valence fermions on MILC \texttt{asqtad} lattices, as well
as Wilson fermions on quenched lattices.  These methods can easily
be extended to $\rho\to\gamma\pi$ transition form factors.

\end{abstract}

\maketitle 

\section{\label{sec:introduction}INTRODUCTION}

The pion electromagnetic form factor is often considered a good observable
for studying the relative accuracy of perturbative and non-perturbative
descriptions of QCD as the energy scale varies.  It is hoped
that because the pion is the lightest and simplest hadron that perturbative
descriptions will remain accurate at lower energy scales than predictions
for heavier and more complicated hadrons, \textit{e.g.}\ the nucleon.

\begin{figure}[ht]
\includegraphics[angle=90,width=0.46\textwidth]{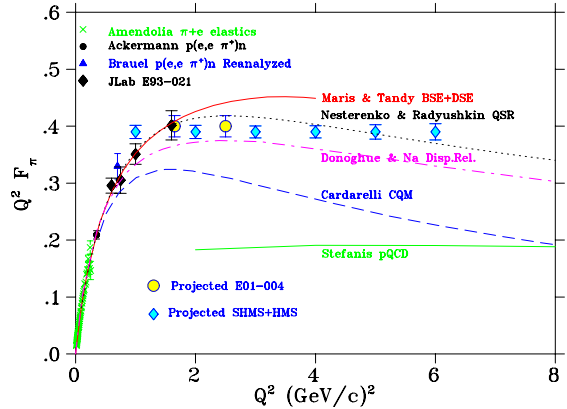}
\vspace{-4ex}
\caption{\label{fig:nucl-ex-0208011-fig-2}Summary of experimental data
and phenomenological predictions for the pion electromagnetic form factor,
taken from \cite{Blok:2002ew}.  Note that some of the points show
projected errors for \textit{proposed} experiments.}
\vspace{-4ex}
\end{figure}

At very low energies, the charged pion moves moves
in an electromagnetic field like a point particle with one unit
of electric charge.  This allows us to fix the overall normalization
of the form factor
\begin{equation}
F_\pi(Q^2) = 1 \quad \mathrm{as} \quad Q^2 \to 0.
\end{equation}
The experimentally observed behavior of the form factor
at small momentum transfer is accurately described
by the vector meson dominance (VMD) hypothesis
\cite{Holladay:1955,Frazer:1959gy,Frazer:1959}
\begin{equation}
\label{eq:vmd_form}
F_\pi(Q^2) \approx \frac{1}{1+Q^2 \left/ m_\rho^2 \right.}
\quad \mathrm{for} \quad Q^2 \ll m_\rho^2
\end{equation}
The current experimental situation is presented
in Fig.~\ref{fig:nucl-ex-0208011-fig-2}, taken from
\cite{Blok:2002ew}.

What is surprising is that VMD with only the lightest $m_\rho$ resonance
appears to describe all the existing data, even up to scales
of $Q^2 \gtrsim 1 \mathrm{GeV}^2$ where one might have hoped
the perturbative QCD (pQCD) would provide accurate predictions.  At very high
momentum transfer, we expect the perturbative description will be correct
as computed in \cite{Brodsky:1973kr,Brodsky:1975vy,Farrar:1979aw}
\begin{equation}
F_\pi(Q^2) = \frac{8\pi\alpha_s(Q^2)f_\pi^2}{Q^2} \quad \mathrm{as} \quad
Q^2 \to \infty
\end{equation}
While the data may indicate an onset of the correct scale dependence
around 2 $\mathrm{GeV}^2$, the numerical value is 100\% larger than
the pQCD asymptotic prediction.

This situation presents many questions.  At what scale does the form factor
scale with $Q^2$ as predicted by pQCD?  Even if we assume proper scaling,
the numerical values do not agree with pQCD.  How rapidly will the data
approach the pQCD prediction and at what scale will pQCD finally agree
with the data?  These are questions which Lattice QCD calculations
are ideally suited to address, provided we can get reliable results
for momentum transfer on the order of a few to several $\mathrm{GeV}^2$.

Early lattice calculations validated the vector meson dominance hypothesis
at low $Q^2$ \cite{Martinelli:1988bh,Draper:1989bp}. Recent lattice results
\cite{vanderHeide:2003ip,Nemoto:2003ng}, including some
of our own preliminary results \cite{Bonnet:2003pf}, have somewhat extended
the range of momentum transfer, up to 2 $\mathrm{GeV}^2$, and the results
remain consistent with VMD and the experimental data.

\section{\label{sec:techniques}LATTICE COMPUTATION OF $F_\pi(Q^2)$}

\begin{figure}[ht]
  \includegraphics[width=0.46\textwidth]{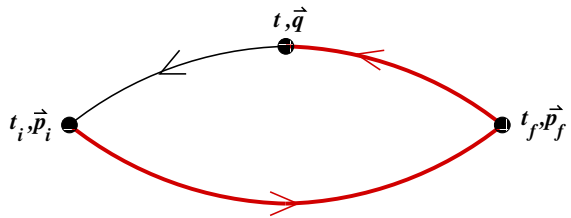}
  \vspace{-4ex}
  \caption{
    \label{fig:threepoint}
    The quark propagators used to compute the pion form factor.
  }
  \vspace{-4ex}
\end{figure}

The electromagnetic form factor is obtained in lattice QCD simulations
by placing a charged pion creation operator at Euclidean time $t_i$,
a charged pion annihilation operator at $t_f$ and a vector current insertion
at $t$ as shown in Fig.~\ref{fig:threepoint}.  A standard quark propagator
calculation provides the two propagator lines that originate from $t_i$.
The remaining quark propagator, originating from $t_f$ is obtained
via the \textit{sequential source method}:  (1) completely specify
the quantum numbers, including momentum $\vec{p}_f$,
of the annihilation operator to be placed at $t_f$ and (2)
contract the propagator from $t_i$ to $t_f$ to the annihilation operator
and use that product as the source vector of a second, sequential
propagator inversion.  The resulting sequential propagator appears
as the thick line in Fig.~\ref{fig:threepoint} extending from $t_i$
to $t$ via $t_f$.  Given these two propagators, the diagram can be computed
for all possible values of insertion position $t$ and source
and insertion momenta, $\vec{p}_i$ and $\vec{q}$,
consistent with overall momentum conservation:
$\vec{q} = \vec{p}_f - \vec{p}_i$.
Furthermore, with the same set of propagators, any current can be inserted
at $t$ and any meson creation operator can be contracted at $t_i$.
So, the diagram relevant to determining the form factor
for the transition $\rho^+ \to \gamma \pi^+$ can be computed without
further quark propagator calculations.  By applying the sequential source
method at the sink, the trade-off is that the entire set of sequential
propagators must be recomputed each time new quantum numbers are needed
at the sink, particularly $\vec{p}_f$.

We can extract the pion energies $E_\pi(\vec{p})$ using standard lattice
techniques of fitting pion correlation functions from which we can compute
the momentum transfer
\begin{equation}
-Q^2 = \left[ E_\pi(\vec{p}_f) - E_\pi(\vec{p}_i) \right]^2
      - ( \vec{p}_f - \vec{p}_i )^2
\end{equation}
which should be non-positive if the pion spectral function is well-behaved.
Since the largest $Q^2$ occur in the Breit frame,
$\vec{p}_f = -\vec{p}_i$, it is important to choose a non-zero $\vec{p}_f$
to achieve large momentum transfer.

The form factor, $F(Q^2)$, is defined by
\begin{eqnarray}
\lefteqn{\left\langle
  \pi(\vec{p}_f) \left| V_\mu(0) \right| \pi(\vec{p}_i)
\right\rangle_\mathrm{continuum} = } \\*
&& Z_V \left\langle
  \pi(\vec{p}_f)\left| V_\mu(0) \right| \pi(\vec{p}_i)
\right\rangle = F(Q^2)(p_i+p_f)_\mu \nonumber
\end{eqnarray}
where $V_\mu(x)$ is the chosen vector current.
The three-point correlation function depicted in Fig.~\ref{fig:threepoint}
is given by
\begin{eqnarray}
\lefteqn{
  \Gamma_{\pi\mu\pi}^{AB}(t_i,t,t_f,\vec{p}_i,\vec{p}_f) =
} \\*
&& a^9 \sum_{\vec{x}_i,\vec{x}_f}
e^{-i(\vec{x}_f-\vec{x})\cdot\vec{p}_f} \ \times
\ e^{-i(\vec{x}-\vec{x}_i)\cdot\vec{p}_i} \nonumber \\*
&& \quad \times \left\langle 0\left|
  \phi_B(x_f) V_\mu(x) \phi_A^\dagger(x_i)
\right| 0 \right\rangle \nonumber
\end{eqnarray}
where $\phi_A^\dagger(x)$ and $\phi_B(x)$ are creation and annihilation
operators with pion quantum numbers.  The $A$ and $B$ indicate that different
operators may be used at the source and sink,
\textit{i.e.}\ smeared source and point sink or pseudoscalar source
and axial vector sink.
 
Inserting complete sets of hadron states and requiring $t_i \ll t \ll t_f$,
gives
\begin{eqnarray}
\lefteqn{
  \Gamma_{\pi\mu\pi}^{AB}(t_i,t,t_f,\vec{p}_i,\vec{p}_f) \to
} \\*
&& a^3 \left\langle 0 \left|
  \phi_B(x)
\right| \pi(\vec{p}_f)\right\rangle
\frac{e^{-(t_f-t)E_\pi(\vec{p}_f)}}{2 E_\pi(\vec{p}_f)} \nonumber \\*
&& \times \left\langle \pi(\vec{p}_f) \left|
  V_\mu(x)
\right| \pi(\vec{p}_i) \right\rangle \nonumber \\*
&& \times \frac{e^{-(t-t_i)E_\pi(\vec{p}_i)}}{2 E_\pi(\vec{p}_i)}
\left\langle \pi(\vec{p}_i) \left|
  \phi_A^\dagger(x)
\right| 0 \right\rangle \nonumber
\end{eqnarray}
Similarly for the two-point correlator and $t_i \ll t_f$
\begin{eqnarray}
\lefteqn{
  \Gamma_{\pi\pi}^{AB}(t_i,t_f,\vec{p}) \to
  a^3 \left\langle 0 \left| \phi_B(x_f) \right| \pi(\vec{p}) \right\rangle
} \\*
&& \times \frac{e^{-(t_f-t_i)E_\pi(\vec{p})}}{2 E_\pi(\vec{p})}
\left\langle \pi(\vec{p}) \left|
  \phi_A^\dagger(x_i)
\right| 0 \right\rangle \nonumber
\end{eqnarray}

For pion operators, we use either the pseudoscalar density,
$\phi^{(1)}(x) = \overline\psi(x) \gamma_5 \psi(x)$, or the temporal
component of the axial vector current,
$\phi^{(2)}(x) = \overline\psi(x) \gamma_5 \gamma_4 \psi(x)$,
at a single point, denoted ``L'' for local, or smeared, denoted ``S,''
using gauge invariant Gaussian smearing 
\begin{equation}
b(x) \to \left(
1 + \frac{\omega}{N} \vec{\nabla} U
\right)^N b(x).
\end{equation}
Note that we have suppressed the flavor structure.  For these operators,
we get the following amplitudes
\begin{eqnarray}
a^2 \left\langle 0 \left| \phi_L^{(1)}(x) \right| \pi(\vec{p}) \right\rangle
& = & Z_L^{(1)} e^{i\vec{x}\cdot\vec{p}} \\
a^2 \left\langle 0 \left| \phi_S^{(1)}(x) \right| \pi(\vec{p}) \right\rangle
& = & Z_S^{(1)}(\left|\vec{p}\right|) e^{i\vec{x}\cdot\vec{p}} \\
a^2 \left\langle 0 \left| \phi_L^{(2)}(x) \right| \pi(\vec{p}) \right\rangle
& = & Z_L^{(2)}(\left|\vec{p}\right|) e^{i\vec{x}\cdot\vec{p}} \\
a^2 \left\langle 0 \left| \phi_S^{(2)}(x) \right| \pi(\vec{p}) \right\rangle
& = & Z_S^{(2)}(\left|\vec{p}\right|) e^{i\vec{x}\cdot\vec{p}}
\end{eqnarray}
We anticipate that $Z_L^{(2)}(\vec{p}) \propto E_\pi(\vec{p})$ based
on the Lorentz structure of the operator.  All of the $Z$'s should
have corrections of $\mathcal{O}(p^2)$.

From this discussion, we can see that there are two ways to determine
the form factor $F_\pi(Q^2)$.  The first method, which we will call
the \textit{fitting method} involves a fit of the relevant two
and three-point functions to simultaneously extract the form factor,
the energies $E_\pi(\vec{p})$ and the amplitudes $Z(\vec{p})$
in a single covariant, jackknifed fit.

Another method, which we will call the \textit{ratio method}
starts by determining the energies $E_\pi(\vec{p})$ and then constructing
the following ratio which is independent of $Z_L^{(1)}$,
$Z_S(|\vec{p}|)$ and all Euclidean time exponentials:
\begin{eqnarray}
F(Q^2,t) &=& \frac{
  \Gamma_{\pi 4 \pi}^{AB}(t_i,t,t_f,\vec{p}_i,\vec{p}_f)
  \Gamma_{\pi\pi}^{CL}(t_i,t,\vec{p}_f)
}{
  \Gamma_{\pi\pi}^{AL}(t_i,t,\vec{p}_i)
  \Gamma_{\pi\pi}^{CB}(t_i,t_f,\vec{p}_f)
} \nonumber \\*
&& \times \left(
  \frac{2 Z_V E_\pi(\vec{p}_f)}{E_\pi(\vec{p}_i+E_\pi(\vec{p}_f)}
\right)
\end{eqnarray}
where the indices $A$, $B$ and $C$ can be either $L$ (local)
or $S$ (smeared).  For a conserved current, $Z_V\equiv1$.
As part of our program, we expect to determine the relative merits
of each extraction method.

\section{\label{sec:simulation}SIMULATION DETAILS}

\begin{table}[t]
  \caption{\label{tab:Wilson_kappas}Simulation details for quenched Wilson
    fermion calculations at $a^{-1} \approx 2\ \mathrm{GeV}$}.
  \begin{tabular}{llll}
    $\kappa$ & volume         & $am_\pi$ & $am_\rho$ \\
    \hline
    0.1480   & $16^3\times32$ & 0.673    & 0.712     \\
    0.1520   & $16^3\times32$ & 0.477    & 0.549     \\
    0.1540   & $16^3\times32$ & 0.364    & 0.468     \\
    0.1555   & $24^3\times32$ & 0.259    & 0.398     \\
    0.1563   & $24^3\times32$ & 0.179    & 0.358     \\
    0.1566   & $24^3\times32$ & 0.145    & 0.343     \\
    0.1566   & $32^3\times48$ & 0.145    & 0.343
  \end{tabular}
  \vspace{-4ex}
\end{table}

Our first calculations were done on quenched configurations generated
with the Wilson gauge action at $\beta=6.0$
($a^{-1} \approx 2\ \mathrm{GeV}$).  The propagators were computed
using the unimproved Wilson fermion action with Dirichlet boundary
conditions.  For Wilson fermions at these lattice spacings, the exceptional
configuration problem is rather mild particularly when compared
to non-perturbatively improved Clover action.  This enabled us to reach
pion masses of 300 MeV without observing any exceptional configurations,
see Tab.~\ref{tab:Wilson_kappas}
whereas Clover simulations are limited to pions of roughly 500 MeV
\cite{vanderHeide:2003ip}.  Of course, pion masses in quenched domain wall
fermion calculations are limited only by finite volume effects and available
computing power.  Yet the best results so far for pion form factors
are 390 MeV pions at $a^{-1} \approx 1.3\ \mathrm{GeV}$ lattice spacing
\cite{Nemoto:2003ng}.  Furthermore, at this workshop we learned that Wilson
fermion results may be automatically $\mathcal{O}(a)$ improved with just
double the effort \cite{Frezzotti:2003xj}, which is an option we have
under consideration
for the future.

\begin{table}[ht]
  \caption{\label{eq:DWF_details}Simulation details for domain wall fermion
    calculations on 272 dynamical MILC \texttt{asqtad} lattices
    at $a^{-1} \approx 1.5\ \mathrm{GeV}$.}
  \begin{tabular}{llllll}
    $am_{u,d}$ & $am_s$ & $am_\mathrm{dwf}$ & Volume & $am_\pi$ \\
    \hline
    0.01       & 0.05   & 0.01 & $20^3\times64$ & 0.200 \\
  \end{tabular}
  \vspace{-4ex}
\end{table}

For our unquenched results, we used valence domain wall fermions
with a domain wall height of $m_0=1.7$ and $L_s = 16$ on MILC
$N_f=2+1$ and $N_f=3$ lattices after HYP blocking
\cite{Hasenfratz:2001hp}.  Dirichlet boundary conditions were imposed
32 timeslices apart for the domain wall propagator calculations.
Detailed results of several quantities computed
on these MILC configurations were presented at this workshop
\cite{Gottlieb:2003bt}.  Preliminary results on related hadronic observables
using many of the same domain wall propagators have recently been
presented \cite{Schroers:2003mf} and were also discussed
at this workshop.

\section{\label{sec:results}PRELIMINARY RESULTS}

\begin{figure*}[thb]
  \begin{center}
    \includegraphics*[height=59mm]{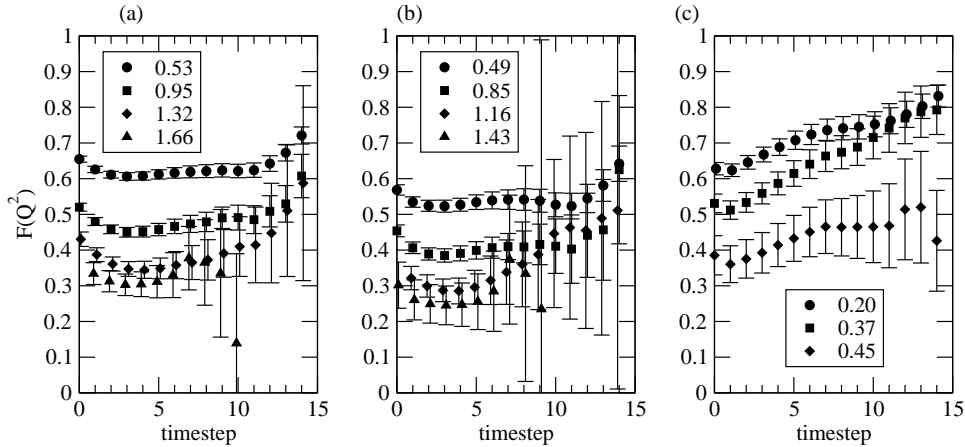}
  \end{center}
  \vspace{-13mm}
  \caption{\label{fig:Wilson_plateaus}Pion form factor data versus timestep,
    for (a) $\kappa$=0.1520, (b) $\kappa$=0.1540 and (c) $\kappa$=0.1563.
    Numerical values of $Q^2$ in GeV are shown in the legends.}
  \vspace{-2mm}
\end{figure*}

Representative plots of the pion form factor and it's dependence
on the timeslice of the current insertion $F_\pi(Q^2, t)$ are
shown in Fig.~\ref{fig:Wilson_plateaus}.  The form factors in these plots
were computed using the ratio method.  For the heavier pion masses
(smaller values of $\kappa$) there are nice plateaus which indicate
reasonable signal to noise for all momenta considered and little
contamination from excited states.  As the pion masses get lighter,
the signal to noise decreases and the plateaus become less convincing
which suggest that the source and sink may not be sufficiently separated
and hence excited states may be contaminating the ratio.
Another possibility is that the ratio may not provide an optimal
determination of the form factor at lighter pion masses.
A detailed study is currently in progress.

\begin{figure*}[hbt]
  \begin{center}
    \includegraphics*[height=59mm]{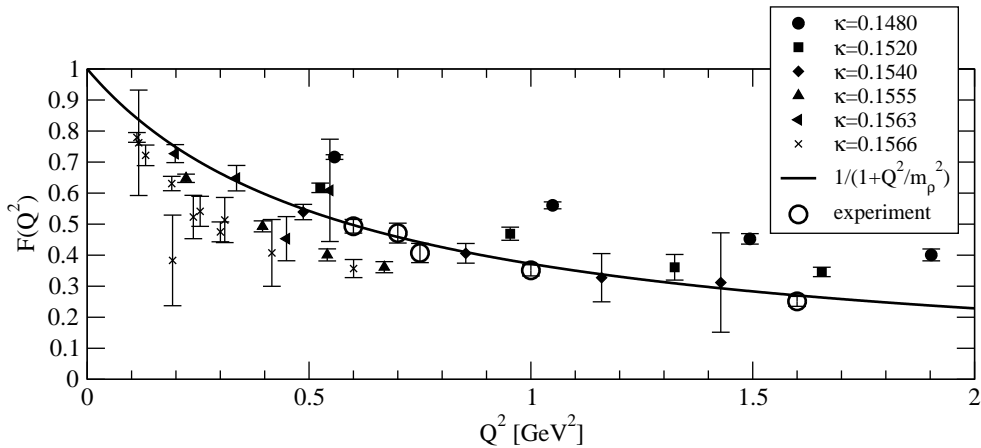}
  \end{center}
  \vspace{-13mm}
  \caption{\label{fig:Wilson_monopole}Results for the pion form factor
    as a function of $Q^2$ for each of the available $\kappa$ values.
    Experimental measurements\protect\cite{Volmer:2000ek}
    and the vector meson dominance hypothesis are also shown.}
\vspace{-2mm}
\end{figure*}

Preliminary results of the quenched Wilson form factor computed
by the fitting method are plotted in Fig.~\ref{fig:Wilson_monopole}.
Also shown in the figure are experimental data points and a curve
showing the prediction of vector meson dominance using the observed
value for the $\rho$ meson mass.  While the data tend in the correct
direction with decreasing pion mass, the reader may notice that the
form factor for 300 MeV pions already lies below the physical curve.
This suggests that fitting the data to extract a lattice determination
of the vector meson mass would underestimate the physical value.
In fact, this is exactly what one should expect since it is known
that $\mathcal{O}(a)$ scaling violations to the vector meson mass
computed with Wilson fermions tend to underestimate the mass
by roughly 20\% \cite{Edwards:1998nh}, the same amount needed to move
the form factor points above the continuum curve.  Hence, the proposal
of Frezzotti and Rossi \cite{Frezzotti:2003xj} becomes all the more appealing
as a straightforward way to remove what is presumably
an $\mathcal{O}(a)$ lattice artifact.

Because we would like to compute the pion form factor at large momentum
transfer, we have spent a substantial amount of effort on our
domain wall data set in extracting the pion energies and amplitudes
at relatively large momenta.  In the continuum limit, the pion
dispersion relation should follow the continuum one
\begin{equation}
E_\pi(\vec{p})^2 = \vec{p}^2 + E_\pi(0)^2
\end{equation}
Another possibility is that the pion dispersion relation
will follow the dispersion relation of a free lattice boson
\begin{eqnarray}
\lefteqn{\frac{1}{4} \sinh^2\left(\frac{E_\pi(\vec{p})}{2}\right) =} \\*
&& \frac{1}{4} \sin^2\left(\frac{\vec{p}}{2}\right)
+ \frac{1}{4} \sinh^2\left(\frac{E_\pi(0)}{2}\right) \nonumber
\end{eqnarray}
Both relations agree in the small momentum limit.

\begin{figure*}
  \vspace{-12ex}
  \begin{center}
    \includegraphics[width=0.45\textwidth]{disp.eps}
    \includegraphics[width=0.45\textwidth]{disp_lat.eps}
  \end{center}
  \vspace{-10ex}
  \caption{\label{fig:dwf_dispersion}Pion dispersion relation
    \textit{vs.}\ continuum (left) and lattice (right) expectations.}
  \vspace{-2ex}
\end{figure*}

In Fig.~\ref{fig:dwf_dispersion}, we have plotted against each
dispersion relation and we see that both dispersion
relations provide a reasonable representation of the data, although
there may be a slight flattening of the data against the continuum curve
at higher momenta.  These results suggest that directly fitting
all the data to either dispersion relation, thereby reducing
the number of fit parameters needed to extract the form factor, may
improve the relative signal to noise of the remaining parameters.
This may help dramatically in the ratio method, where the only fit parameters
are the form factor and the energies.

\begin{figure*}
  \vspace{-9ex}
  \begin{center}
    \begin{minipage}{0.45\textwidth}
      \includegraphics[width=0.95\textwidth]{zs1.eps}
    \end{minipage}
    \begin{minipage}{0.45\textwidth}
      \includegraphics[width=0.95\textwidth]{zl1.eps}
    \end{minipage} \\
    \vspace{-6ex}
    \begin{minipage}{0.45\textwidth}
      \includegraphics[width=0.95\textwidth]{zs2.eps}
    \end{minipage}
    \begin{minipage}{0.45\textwidth}
      \includegraphics[width=0.95\textwidth]{zl2.eps}
    \end{minipage}
  \end{center}
  \vspace{-9ex}
  \caption{\label{fig:dwf_amplitudes}Results for the four pion source
    amplitudes $Z_A^{(i)}(\vec{p})$ used in this study.}
  \vspace{-2ex}
\end{figure*}

In the fitting method, one must reliably estimate not only the energies,
but the amplitudes, at high momenta. In Fig.~\ref{fig:dwf_amplitudes}
we present the four amplitudes we estimate from the four two-point
correlators we measure: smeared-smeared and smeared-local for both
pseudoscalar-pseudoscalar and axial-axial operators.  In the fitting
procedure, all four correlators are constrained to have the same energy.
From the figure, we can see that our expectations
of $Z_L^{(1)} \propto \mathrm{const}$
and $Z_L^{(2)}(\vec{p}) \propto E_\pi(\vec{p})$ are consistent
with the data.  We can also see from $Z_S^{(1)}$ that the smeared
pseudoscalar operator has a strong overlap with the zero momentum pion
but that the overlap diminishes rapidly with increasing momenta.
From $Z_S^{(2)}$ we see that the smeared axial-vector operator
has good overlap over a wide range of non-zero momenta, with maximal overlap
around $ap \sim 0.5$.  Thus, given the variety of sources and their relative
strengths of overlapping with pions at different momenta was key
to fitting pion energies up to such high momenta.

\section{\label{sec:Conclusions}CONCLUSIONS}

From our preliminary quenched Wilson form factor results, we find
that both the ratio method and the fitting method are useful tools
for computing the pion form factor.  Each method has different systematic
errors, so the extent to which both agree should give confidence
that the systematic errors are small and well understood.

From our preliminary dynamical domain wall spectrum results, we recognize
the importance of using a large basis of pion operators so that at least
one will have reasonable overlap with the momenta under consideration.
The local axial vector operator is particularly useful in fitting
higher momentum states since it's overlap with a given state increases
as the energy increases.

In the near future, when the pion form factor analysis is complete,
it will be a straightforward extension to our existing analysis framework
to compute the $\rho^+\to\gamma\pi^+$ transition form factors
using the propagators and sequential propagators we have already computed.

\bibliography{main}

\end{document}